\documentclass[manuscript]{emulateapj}
\usepackage{natbib}
\usepackage{lineno}
\usepackage{multirow}
\usepackage{booktabs}
\usepackage{epsfig}
\usepackage{ulem}
\usepackage{amsmath}
\usepackage{soul}
\usepackage[dvipsnames]{xcolor}

\usepackage{amssymb}
\usepackage{amsmath}
\usepackage{graphicx}
\usepackage{hyperref}
\usepackage{mathtools}

\usepackage{savesym}
\savesymbol{tablenum}
\usepackage{siunitx}
\restoresymbol{SIX}{tablenum}

\usepackage{gensymb}

\newcommand{\urlcode}{https://github.com/djones1040/PS1_surface_brightness}
\newcommand{\urldessn}{https://www.darkenergysurvey.org/des-year-3-supernova-cosmology-results/}
\newcommand{\urldr}{https://panstarrs.stsci.edu}

\newcommand{\col}{{\cal C}}
\newcommand{\omegam}{{\Omega_{m}}}
\newcommand{\omegal}{{\Omega_{\Lambda}}}
\newcommand{\G}{{G_{\text{host}}}}
\newcommand{\mubias}{{\Delta \mu_{\text{bias}}}}
\newcommand{\FSB}{F_{\rm SB}}

\newcommand{\ntotsn}{292}
\newcommand{\ntotf}{172}
\newcommand{\ntotl}{120}
\newcommand{\ndessn}{207}

\shorttitle{Probing systematic bias in low-redshift type Ia supernova measurements by cross-analyzing SB and HR}

\begin{document}

\title{Probing Systematic Bias in Low-Redshift Type Ia Supernova Measurements by Cross Analyzing Surface Brightness and Hubble Residuals}

\def\andname{}

\author{                                                H.~Solak\altaffilmark{1},                       R.~Kessler\altaffilmark{2,3},                                           D.~O.~Jones\altaffilmark{4,5}
}

\affil{$^{1}$ Department of Physics, University of Chicago, Chicago, IL 60637, USA}
\affil{$^{2}$ Department of Astronomy and Astrophysics, University of Chicago, Chicago, IL 60637, USA}
\affil{$^{3}$ Kavli Institute for Cosmological Physics, University of Chicago, Chicago, IL 60637, USA}
\affil{$^{4}$ Department of Astronomy and Astrophysics, University of California, Santa Cruz, CA 92064, USA}
\affil{$^{5}$ NASA Einstein Fellow}

\submitted{Accepted by PASP}

\begin{abstract}
For low-redshift $(z < 0.1)$ Type Ia supernovae (SN~Ia) samples used in several cosmological analyses over the past decade, we probe for systematic bias by looking for correlations between surface brightness (SB) measurements and Hubble residuals (HR). For {\ntotsn} SNe~Ia, we measure SB at the location of the SN~Ia from publicly available Pan-STARRS (PS1) images. The Hubble residuals are from two recent measurements with low-$z$ SNe~Ia that overlap the PS1 footprint: 1) the DES 3-year cosmology analysis, with {\ntotl} overlapping low-$z$ SNe~Ia from the Harvard-Smithsonian Center for Astrophysics surveys and Carnegie Supernova Project, and 2) the PS1 single-telescope analysis, with {\ntotf} overlapping low-$z$ SNe~Ia from the Foundation Supernova Survey. This study is motivated by previous reports of anomalous inefficiencies and flux scatter for transients on bright galaxies. We compare HR distributions of the bright and faint halves of the SB distribution: the mean HR values differ by $\Delta \overline{\text{HR}}$ = $0.031 \pm 0.018$, consistent with no difference at the 2$\sigma$ level. We also perform a Kolmogorov-Smirnov (KS) test for the bright and faint half HR distributions, and conclude that the two distributions are statistically consistent with a KS p-value of 0.07. However, if future studies with larger datasets find $\Delta \overline{\text{HR}} \sim 0.03$ with high significance, this difference would be a leading systematic uncertainty in measurements of the dark energy equation of state, $w$.
\end{abstract}

\section{Introduction}
\label{sect:intro}

Since the discovery of cosmic acceleration (\citealt{Riess98}, \citealt{Perlmutter99}), Type Ia supernova (SN~Ia) distance measurements continue to be a critical analysis tool for measuring cosmic distances and the properties of dark energy. To measure the dark energy equation of state, $w$, with high precision, accurate SN~Ia photometry and calibration is crucial. Efforts to improve these measurements have been made by large-area surveys such as the Sloan Digital Sky Survey (SDSS; \citealt{sdss}), Supernova Legacy Survey, (SNLS; \citealt{snls1}, \citealt{snls2}), Pan-STARRS (PS1; \citealt{pantheon}, \citealt{Jones19}), Dark Energy Survey (DES; \citealt{DESkp}), and Joint Light-curve Analysis (JLA; \citealt{Betoule14}). These analyses find that cosmic acceleration is consistent with a cosmological constant ($w = -1$), and the combined statistical and systematic uncertainty on $w$ is $\sim$0.05. As part of these measurements, it is important to quantify systematic uncertainties from many effects including zero-points, filter transmissions, and spectral energy distribution (SED) dependencies. Here we explore a new systematic effect related to the local surface brightness (SB) of the underlying galaxy.

This search is motivated by recent reports of anomalous detection inefficiencies and flux scatter of transients on bright galaxies, where anomalous refers to effects that are much larger than expectations from increased Poisson noise. For the DES transient detection pipeline, \citet{Kessler15} used fake SN~Ia light curves overlaid on images to show that flux uncertainties are underestimated in proportion to local galaxy SB (see Fig.~10 of \citealt{Kessler15}). For the brightest sources, which generally correspond to lower redshifts ($z \sim 0.1$), the SN~Ia flux uncertainty is under-estimated by about a factor of 5. For a kilonova search using data from DES, \citet{Doctor17} also examined fake transients and found decreasing detection efficiency for faint objects with higher underlying SB (see Fig.~7 of \citealp{Doctor17}). The explanation for this effect is not known but hypotheses include errors in point-spread function (PSF) modelling, atmospheric refraction, and effects from pixel correlation.

These issues raise concerns about the existence of data reduction artifacts in the photometry of low-redshift SNe~Ia. Artifacts such as SB-related biases have not been explored in low-redshift SN~Ia samples because the underlying local SB information was not provided in public data releases.

Understanding SB artifacts is important not only to correct for distance biases, but also because many studies have found a relationship between Hubble residuals (HR) and the properties of host galaxies. The most commonly used relationship is a $\sim0.05$~mag step of HR versus stellar mass \citep{Kelly10, Sullivan10, Lampeitl10}.
Relationships between HR and galaxy properties at the SN location have also been found using color, specific star formation rate, and mass \citep{Rigault13,Jones18,Rigault18,Roman18,Kim19,Kelsey20}. However, these measured correlations could be partially caused by data reduction artifacts, and such artifacts are expected to leave signatures in a SB-HR plot.

Here we study the correlation between HR and SB. We use the publicly available PS1 images, covering 3$\pi$ of the sky, to measure the local host galaxy SB at the location of {\ntotsn} low-redshift SN~Ia.
The code for making these SB measurements is available online\footnote{\url{\urlcode}}. The Hubble residuals (HR) are from both the DES 3-year (DES-SN3YR) cosmology analysis \citep{DESkp} and \citet{Jones19}. To illustrate the $w$-sensitivity in the DES-SN3YR analysis, a 0.01~mag bias in the average low-redshift HR results in a $w$-bias of $\sim$0.02, which would be one of the dominant systematic errors in the analysis.

The outline of this letter is as follows.
In  \S\ref{sect:data}, we describe the low-redshift SNe~Ia sample and PS1 imaging data.
In  \S\ref{sect:analysis}, we describe the DES 3-year cosmology analysis and our method of measuring local SB using PS1 images.  
In  \S\ref{sect:results}, we present our correlation study between SB and HR. 
In  \S\ref{sect:conclusion}, we state our conclusions and give suggestions for future work.

\section{Data Sample}
\label{sect:data}

\subsection{The low-redshift SN~Ia sample}

We use a low-redshift $(0.005 < z < 0.110)$ SN~Ia sample that includes the Harvard-Smithsonian Center for Astrophysics surveys \citep[CfA3, CfA4;][]{CfA3,CfA4}, the Carnegie Supernova Project \citep[CSP;][]{CSP, CSP2}, and Foundation \citep{Foley18}. The CfA, CSP sample of {\ntotl} SNe~Ia has been used in several cosmology analyses over the past decade, including the JLA \citep{Betoule14}, Pantheon \citep{pantheon}, and DES-SN3YR \citep{DESkp}. The Foundation sample includes {\ntotf} low-redshift SNe~Ia that satisfy the criteria in \citet{Foley18}; this sample was used to measure cosmological parameters in \citet{Jones19}. To extract PSF fitted photometry for the SNe, CfA and Foundation used DoPHOT \citep{dophot} and CSP used DAOPHOT \citep{daophot}.

The low-$z$ SN~Ia sample used in this analysis includes: 72 from CfA3, 38 from CfA4, 10 from CSP, 172 from Foundation. The DES-SN3YR analysis includes 2 additional low-$z$ SNe~Ia (122 total) that are outside the PS1 footprint. CfA1 and CfA2 are not included because DES-SN3YR only uses events with measured telescope+filter transmissions. 


\subsection{PS1 imaging data}

PS1 is a 1.8-meter telescope with a 1.4-gigapixel camera (GPC1; \citealt{Waters16}). PS1 utilizes a wide-field astronomical imaging and data processing facility developed and operated by the Institute for Astronomy at the University of Hawaii \citep{Kaiser10}. GPC1 has a pixel size of $10~\mu$m which subtends \SI{0.258}{\arcsecond} and is well below the FWHM size of the point spread function (PSF) of $\sim$\SI{1.3}{\arcsecond} \citep{Chambers16}.

All PS1 images are processed through the Image Processing Pipeline (IPP) at the Maui High Performance Computer Center. The pipeline processes the images through a series of stages, including de-trending or removing the instrumental signature, a flux-conserving warping to a sky-based image plane, masking and artifact removal, object detection and photometry, and sky-subtraction \citep{Chambers16}.

The data we use to make SB measurements is from PS1 3$\pi$ Steradian Survey, publicly available from the PS1 data release 2 (DR2)\footnote{\url{\urldr}}. DR2 covers the entire sky above declination $-30\degree$. The dataset contains stacked, sky subtracted images, and has uniform calibration to within 0.005 mag \citep{Schlafly12}. For the $gri$ filters used in our SB analysis, the mean 5$\sigma$ point source limiting sensitivities are 23.3, 23.2, 23.1 mag, respectively.

\section{Analysis}
\label{sect:analysis}

\subsection{Low-redshift Hubble residuals}
The luminosity distance ($d_L$) dependence on cosmological parameters is
\begin{equation}
    d_{L} = (1+z)c\int_{0}^{z}\frac{dz'}{H(z')}~,
\end{equation}
where 
\begin{equation}
    H(z) = H_{0}[\omegam(1+z)^{3} + \omegal(1+z)^{3(1+w)}]^{1/2}~,
\end{equation}
and the Hubble constant ($H_0$), matter density ($\omegam$), and dark energy density ($\omegal$) are defined at redshift 0. The dark energy equation of state parameter is $w$, and $w = -1$ for a cosmological constant. The $\Lambda$CDM model distance modulus is defined as $\mu_{\text{model}} = 5\log(d_{L}/10\text{pc})$.

As part of measuring $\mu$ in the DES-SN3YR and Foundation analyses, the standardization of SNe~Ia is based on color and stretch parameters determined from a light curve fit. For each SN~Ia they used the SALT2 model from \citet{Betoule14} to determine the amplitude ($x_{0}$), light curve width ($x_{1}$), and color ($\col$). These SALT2 parameters were used to measure the distance modulus using a modified Tripp equation \citep{Tripp98}:
\begin{equation}
    \mu = m_{B} + \alpha x_{1} - \beta C + M_{0} + \gamma
    \G + \mubias~,
    \label{eq:mu}
\end{equation}
where $m_{B} = -2.5\log(x_{0})$, $\alpha$ and $\beta$ are the nuisance parameters describing the brightness-stretch and brightness-color relations, $M_0$ is the absolute SN~Ia magnitude with $\col = x_1 = 0$ ($H_0 = 70 {\rm~km~s^{-1}~Mpc^{-1}}$), $\gamma \G$ is the dependence of the shape- and color-corrected SN magnitude on host galaxy stellar mass \citep{Conley11}, and $\mubias$ is a bias correction determined from simulations \citep{Kessler19, Jones19}. See \citet{DESkp} Eq.~4 for more details. For this sample, the nuisance parameters $(\alpha,\beta,\gamma,M_0)$ were determined using the methodology from ``BEAMS with Bias Corrections'' (BBC; \citealt{bbc}).

The best fit cosmological model was obtained using CosmoMC \citep{cosmomc} which uses SN~Ia distances (Eq.~\ref{eq:mu}) and a prior from the cosmic microwave background (CMB: \citealt{planck16}). From the CosmoMC fit the Hubble residuals are defined as 
\begin{equation}
    \text{HR} = \mu - \mu_{\text{model}}.
\end{equation}
We do not repeat this analysis but instead we take the HR values from the public data releases for DES-SN3YR\footnote{\url{https://des.ncsa.illinois.edu/releases/sn}} and Foundation\footnote{\url{https://github.com/djones1040/Foundation_DR1}}. 


\subsection{SB measurements using PS1 images}

To make SB measurements at SN~Ia locations, we use a circular aperture with a radius of \SI{1}{\arcsecond} on PS1 images from DR2 and extract calibrated flux measurements in the $gri$ filters as well as uncertainty values. The choice of radius comes from the $\sim$\SI{1}{\arcsecond} FWHM of the PS1 PSF. The pixel flux contributing to the SB flux is calculated using $f = F\cdot F_A$, where $f$ is the contributing pixel flux, $F$ is the total pixel flux, and $F_A$ is the fraction of each pixel contained inside the circle. The SB flux is defined as $ \FSB = [\sum_i f_i ] / \pi R^2$, where $i$, $R=\SI{1}{\arcsecond}$, and PS1 fluxes are scaled to match the DES zero-point.

\begin{figure}[h]
    \begin{centering}
    \includegraphics[scale=0.4]{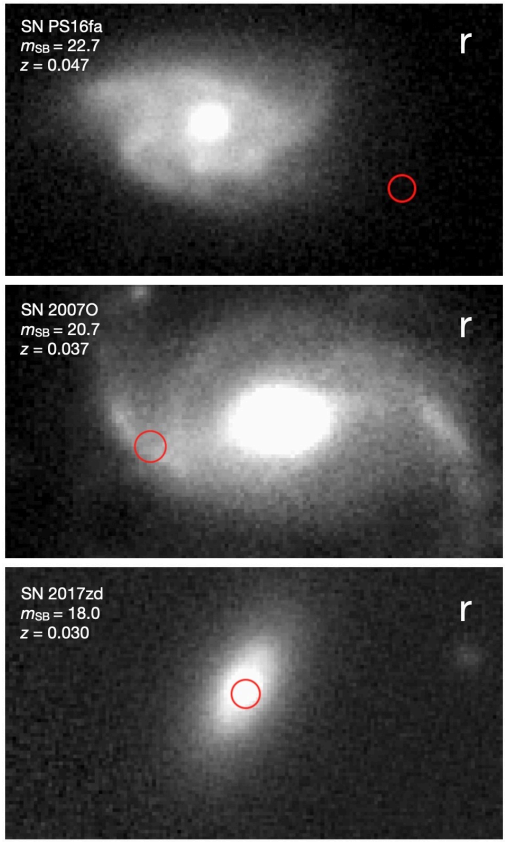}
    \caption{PS1 image stamps of three low-$z$ SN~Ia host galaxies with surface brightness magnitudes ($m_{\text{SB}}$) indicated on each panel. The circle is at the location of the SN~Ia and indicates the aperture used to measure $m_{\text{SB}}$.}
    \label{fig:galaxy}
    \end{centering}
\end{figure}

Fig.~\ref{fig:galaxy} shows PS1 image stamps of three low-$z$ SN~Ia host galaxies with varying surface brightness magnitudes ($m_{\text{SB}}$) and their circular apertures.

\subsection{Cross-check with DES}

Before using SB measurements of the low-redshift sample, we perform a cross-check using the DES subset of the DES-SN3YR sample that includes the $gri$ SB measurements in their data release. The  DES-SN  sample  was  collected over three 5-month-long seasons, from August 2013 to February 2016, using the Dark Energy Camera (DECam, \citealt{Flaugher15}) at the Cerro Tololo Inter-American Observatory. New transients were discovered using a difference-imaging pipeline \citep{Kessler15}. These SB measurements are made on deep coadded templates using images with the best seeing.

Out of {\ndessn} DES SN~Ia events, images were retrieved for 64, 100, and 111 SNe~Ia that overlap with the PS1 footprint and have $F_{SB(PS1)}, F_{SB(DES)} > 10$ ($m<25$) for the $gri$ filters, respectively. Here we compare independent SB measurements from DES and PS1.

\begin{figure}[h]
    \begin{centering}
    \includegraphics[scale=0.55]{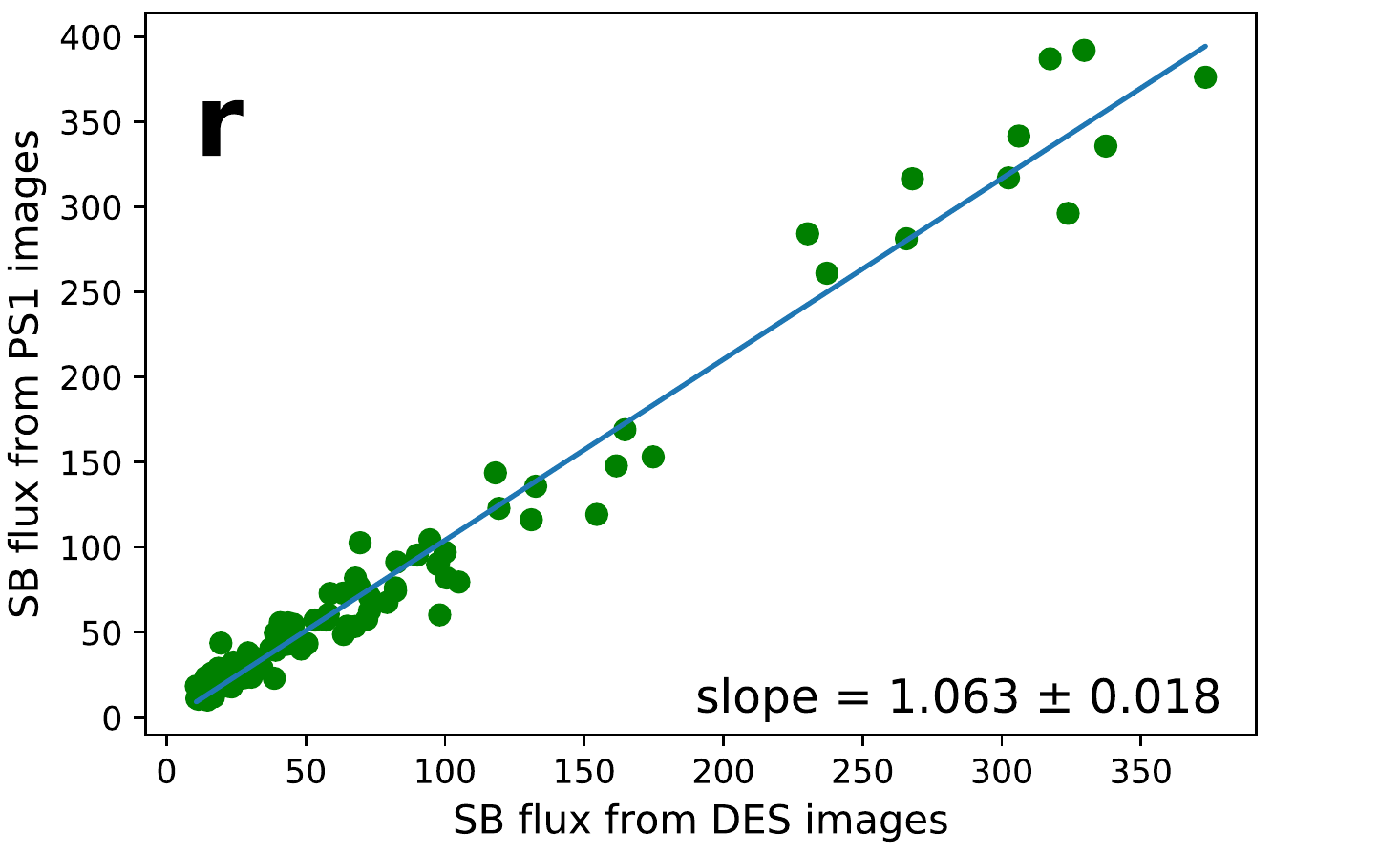}
    \caption{$\FSB$ measurements available from public DES-SN3YR data is compared with our PS1 measurements.}
    \label{fig:cross}
    \end{centering}
\end{figure}

We converted the PS1 SB measurements to the same zero-point as DES. The $gri$ filters for DES and PS1 are similar; the difference between the mean wavelengths $(\bar{\lambda}_{PS1} - \bar{\lambda}_{DES})$ of the $gri$ filter responses are \SI{39}{\angstrom}, \SI{-220}{\angstrom}, and \SI{-284}{\angstrom}, respectively. For this cross-check, we do not K-correct SB measurements to account for filter differences. To estimate the error in the ratio of measured SB flux values ($F_{SB(PS1)}/F_{SB(DES)}$) we use six galaxy spectra from \citet{Coleman80} and \citet{Kinney96} and apply the PS1 and DES filter transmissions to the galaxy SEDs. The mean differences for the $gri$ filters are $2.4\%$, $6.9\%$, and $11.6\%$, respectively. These PS1-DES differences are approximate uncertainties because we did not use spectra from the SN~Ia host galaxies. To estimate the effect of different PSF sizes, we compare SB measurements between DES DR1 data \citep{DESDR1} and PS1 DR2 data using the same algorithm and found systematic SB differences of up to 30\%.


We make a linear fit to $F_{SB(PS1)}$ vs.\ $F_{SB(DES)}$ for each band: the slopes are $1.219 \pm 0.024, 1.063 \pm 0.018$, and $1.130 \pm 0.016$ for the $gri$ bands, respectively. Fig.~\ref{fig:cross} shows $F_{SB(PS1)}$ versus $F_{SB(DES)}$ for the $r$ band. Including the filter transmission and PSF uncertainties, the observed slopes are consistent with 1.0. Considering the filter and PSF size differences between PS1 and DES, the cross-check validates our SB measurement method.



\section{Results}
\label{sect:results}

\begin{figure}[h]
    \begin{centering}
    \includegraphics[trim= 0 0.6cm 0 0, clip=true, scale=0.55]{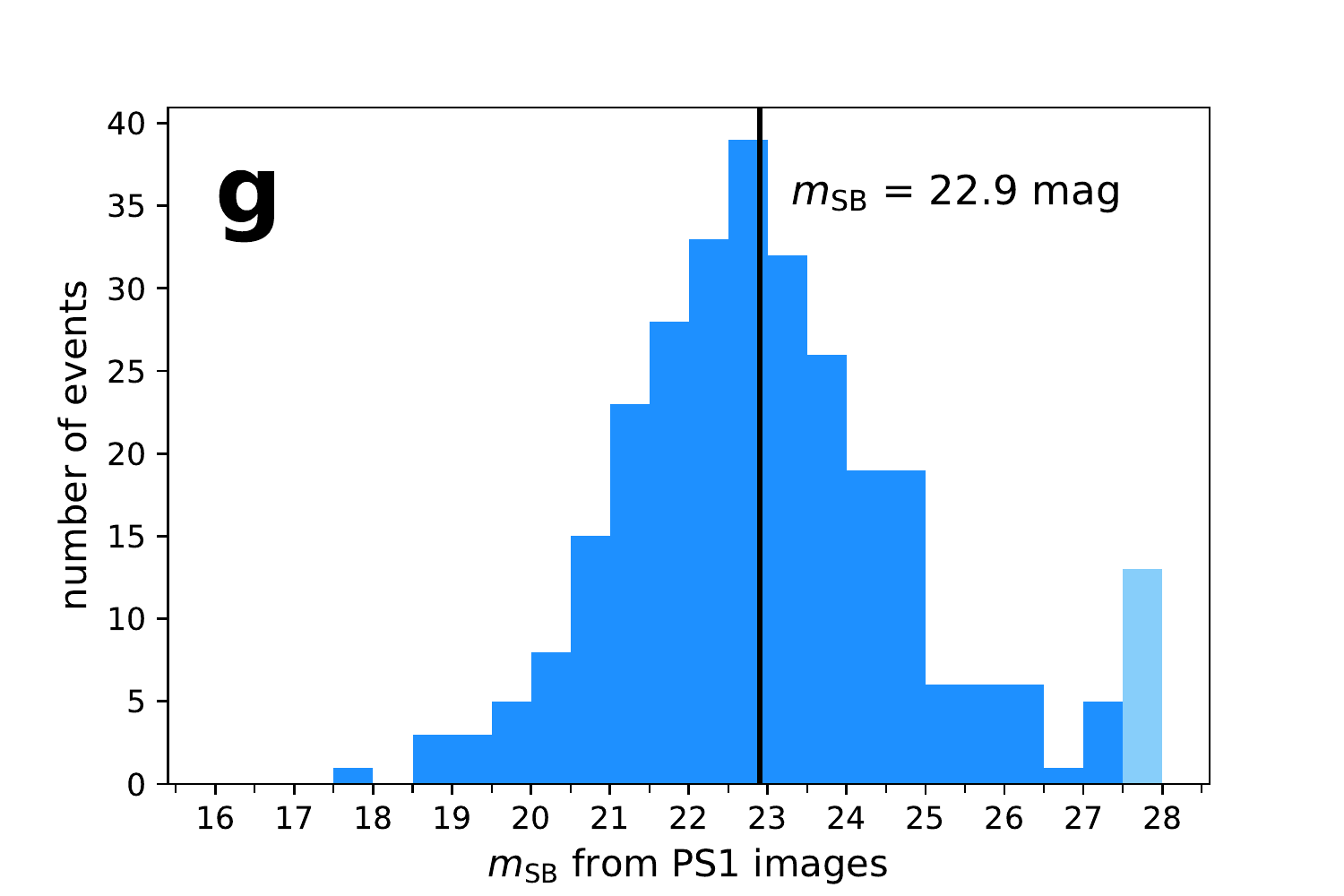}
    \includegraphics[trim= 0 0.6cm 0 0, clip=true, scale=0.55]{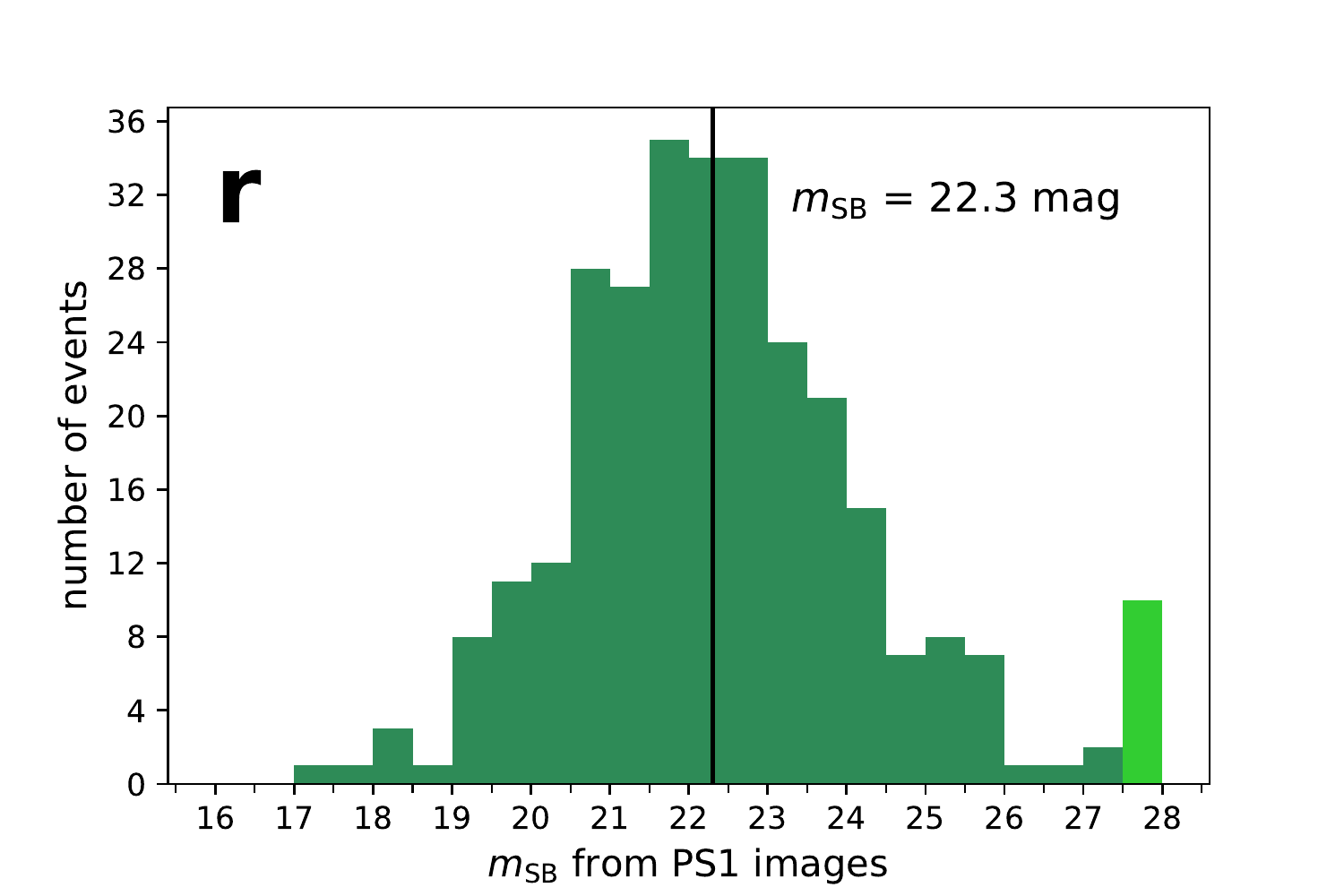}
    \includegraphics[scale=0.55]{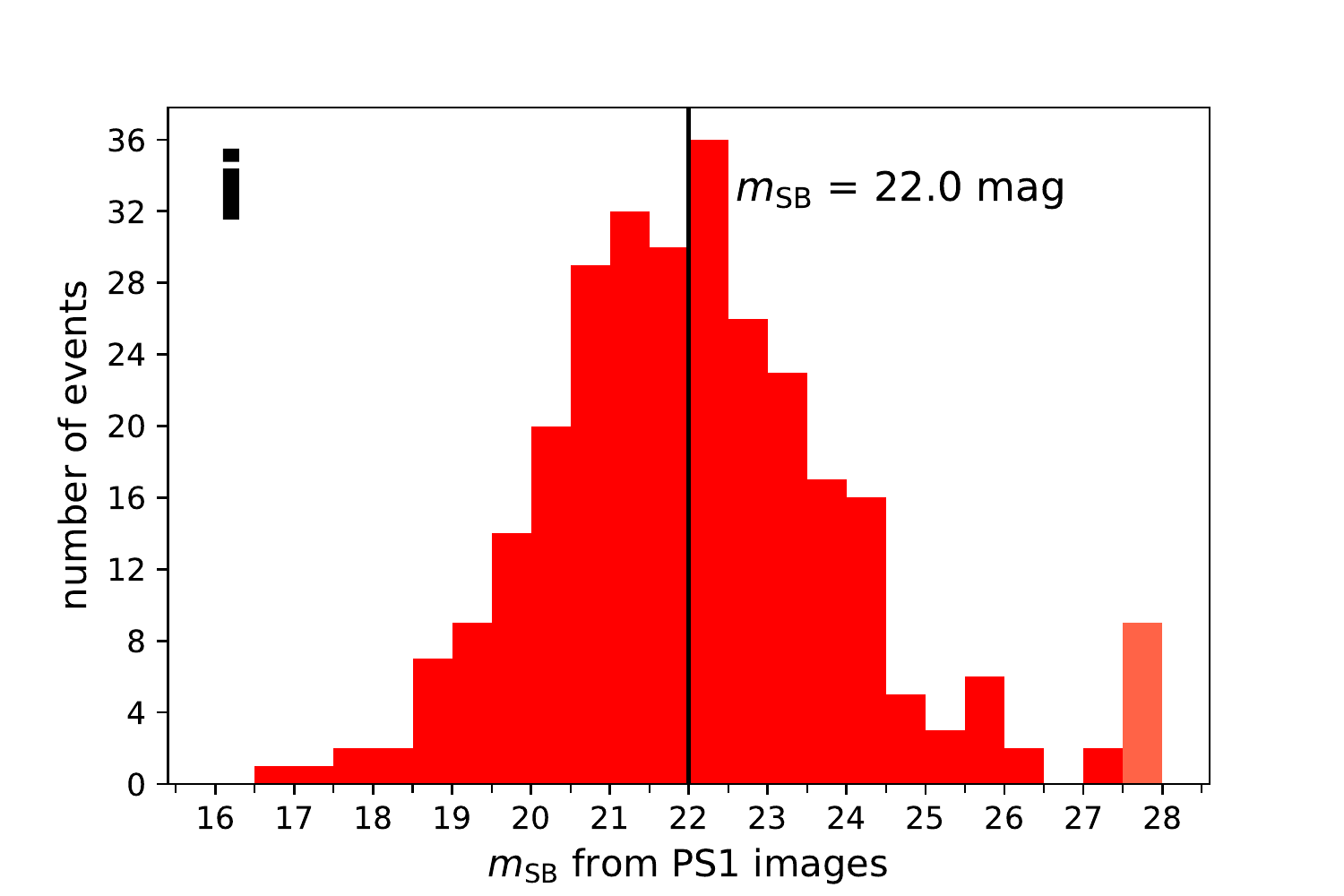}
    \caption{$m_{\text{SB}}$ distributions of the low-redshift sample in $gri$ filters. Vertical line indicates median value. The overflow bin (27.5-28.0) includes events with $\FSB < 0$.}
    \label{fig:calflux}
    \end{centering}
\end{figure}

\begin{figure}[h]
    \begin{centering}
    \includegraphics[scale=0.55]{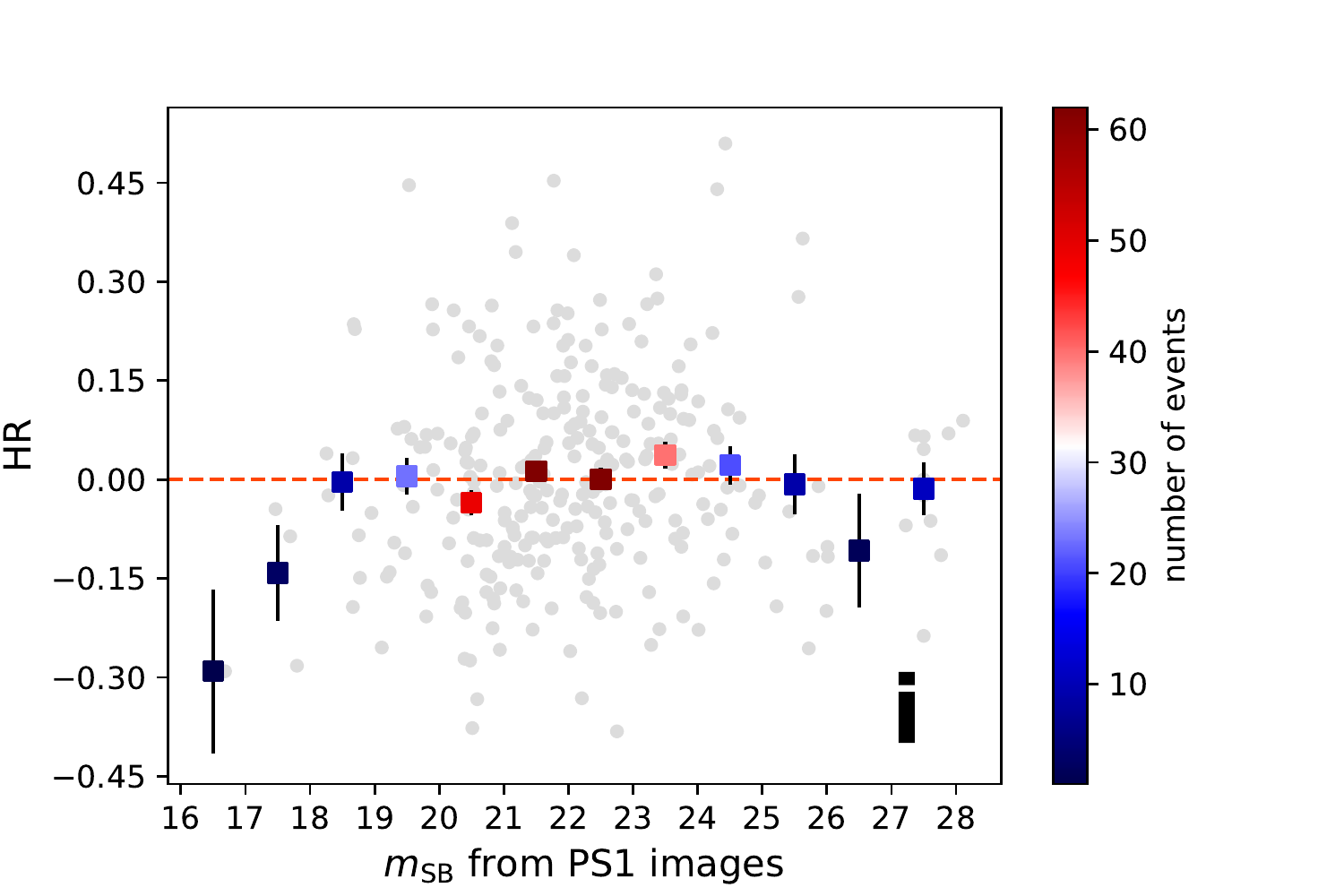}
    \caption{Weighted HR vs.\ $i$-band $m_{\text{SB}}$ with 1 mag bin size. The bins with $m_{\text{SB}} = 16\text{-}17$ and $m_{\text{SB}} = 26\text{-}27$ contain one event.}
    \label{fig:hrsb}
    \end{centering}
\end{figure}

In total we measured surface brightness values for {\ntotsn} low-redshift SNe~Ia for which we have Hubble residuals from the DES-SN3YR and PS1 cosmology analyses. Fig.~\ref{fig:calflux} shows SB magnitude $(m_{\text{SB}})$ distributions in the $gri$ filters. The vertical line indicates median $m_{\text{SB}}$. The left and right sides of the vertical line are defined as the bright half and faint half, respectively. Table~1 shows the number of events in the bright and faint halves for each low-redshift sub-sample and for each band. While the total number of bright and faint $m_{\text{SB}}$ events is the same by definition, all bright and faint sub-sample sizes are consistent as well. Fig.~\ref{fig:hrsb} shows inverse-variance weighted HR vs.\ $i$-band $m_{\text{SB}}$. The residuals are consistent with zero (reduced $\chi^2 = 1.7$), although there is a hint of a bias in the brightest $m_{\text{SB}}$ bins. Using the $g$ and $r$ bands, we find similar results with reduced $\chi^2 = 2.0 \text{ and } 1.6$, respectively.
\vspace*{-5mm}
\begin{table}[h]
\begin{center}
\caption{Low-$z$ sample statistics}
\begin{tabular}{ |c|c|c|c|c| }
\hline
&&&&\\[-2ex]
 & & $N_{\text{evt}}$\tablenotemark{a} & $N_{\text{evt}}$ & \\
filter & survey & bright half & faint half & total\\
\hline
&&&&\\[-2ex]
\multirow{4}{*}{$g$} &CfA3 & 42  & 30 & 72 \\
 &CfA4 & 17 & 21 & 38 \\ 
 &CSP & 6 & 4  & 10\\
 &Foundation & 81 & 90 & 171\tablenotemark{b} \\
\hline
&&&&\\[-2ex]
\multirow{4}{*}{$r$} &CfA3 & 43 & 29 & 72 \\
 &CfA4 & 18 & 20 & 38\\ 
 &CSP & 5 & 5 & 10 \\
 &Foundation & 80 & 91 & 171\tablenotemark{c}\\
\hline
&&&&\\[-2ex]
\multirow{4}{*}{$i$} &CfA3 & 42 & 30 & 72 \\
 &CfA4 & 18 & 20 & 38 \\ 
 &CSP & 5 & 5 & 10 \\
 &Foundation & 81 & 91 & 172 \\
\hline
\end{tabular}
\end{center}
$^{{\rm a}}$ Number of SNe~Ia\\
$^{{\rm b}}$ One Foundation event (SN ATLAS16dqf) was removed for $g$ filter due to image corruption within aperture\\
$^{{\rm c}}$ One Foundation event (SN ASASSN-15uw) was removed for $r$ filter due to nonexistent image.
\label{table:samplestats}
\end{table}

\begin{figure}[h]
    \begin{centering}
    \includegraphics[trim= 0 0.6cm 0 0, clip=true, scale=0.55]{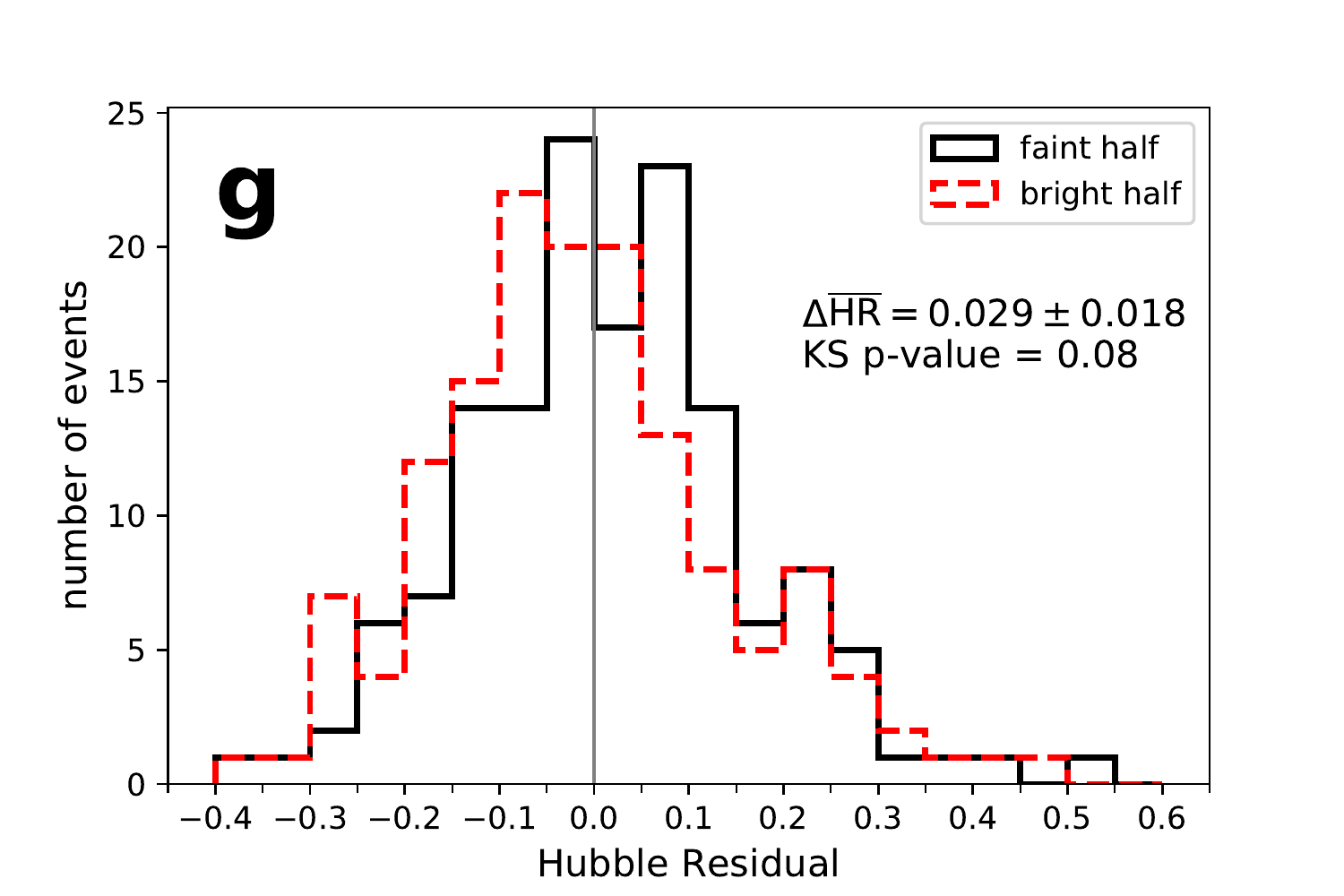}
    \includegraphics[trim= 0 0.6cm 0 0, clip=true, scale=0.55]{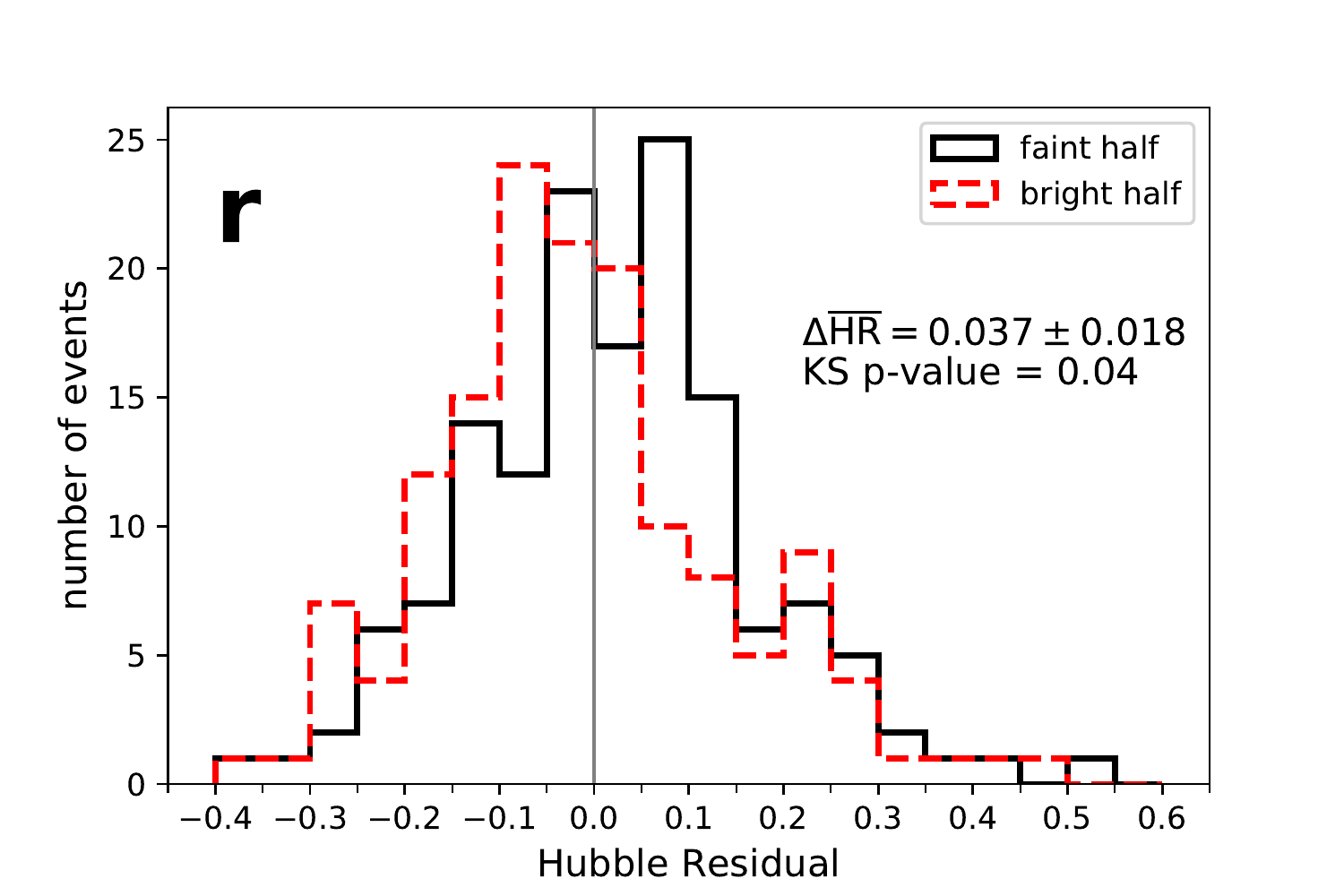}
    \includegraphics[scale=0.55]{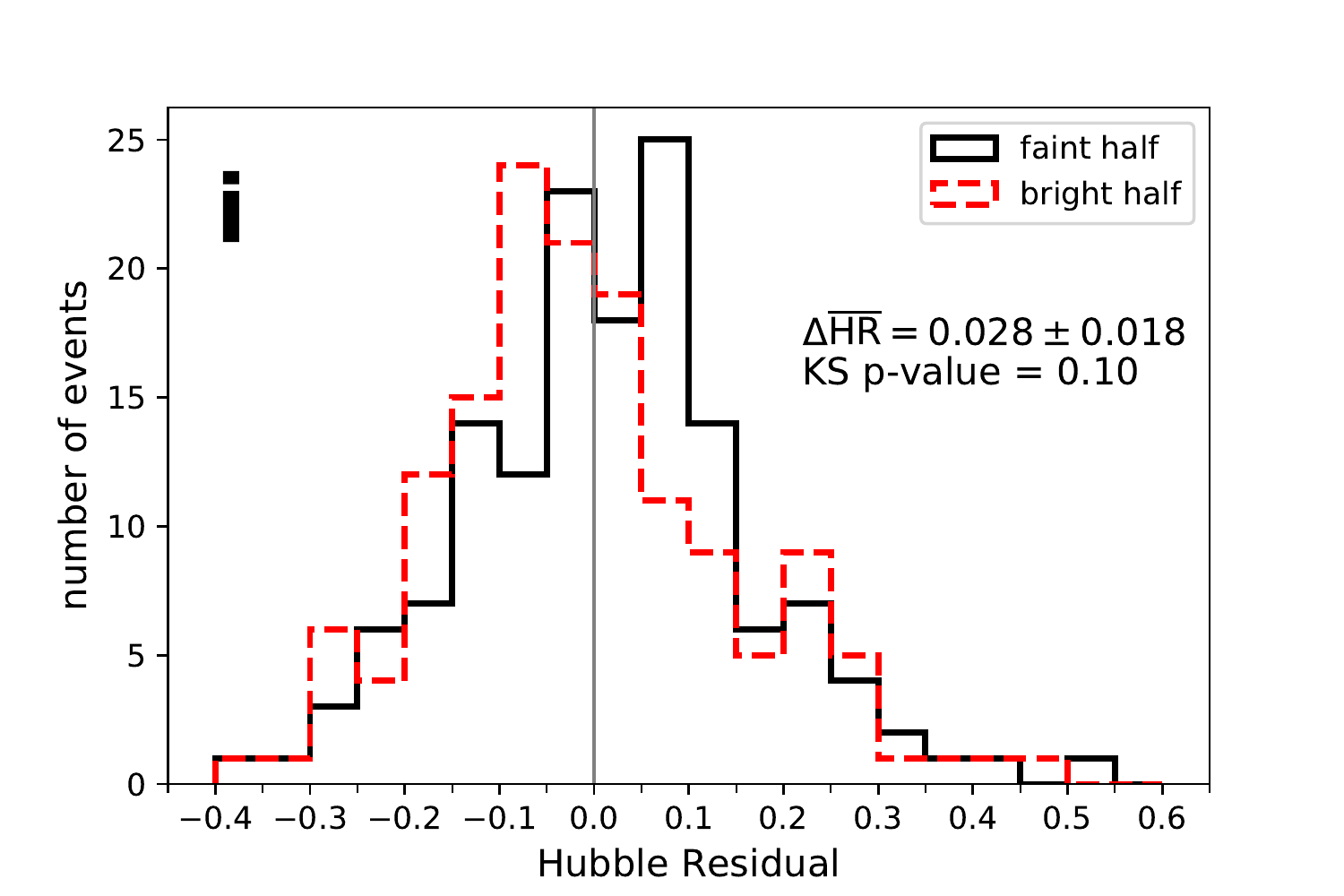}
    \caption{Superimposed HR distributions of the faint (black solid) and bright (red dashed) sub-samples for the $gri$ filters. Difference between the means ($\Delta \overline{\text{HR}}$) and the KS p-value are shown on each panel.}
    \label{fig:supimp}
    \end{centering}
\end{figure}

Next, we compare the bright and faint half HR distributions with a two-sampled Kolmogorov-Smirnov (KS) test. Fig.~\ref{fig:supimp} shows overlaid HR distributions of the bright and faint sub-samples and the results of the comparison. Table~2 shows the difference between the mean HR values ($\Delta \overline{\text{HR}}$) and rms ratios of the bright and faint half distributions, along with KS p-values. For each band, $\Delta \overline{\text{HR}}$ are consistent with zero at the $2\sigma$ level and the rms ratios are consistent with 1 at the $1\sigma$ level. The KS p-values are $0.08, 0.04, 0.10$ for the $gri$ bands, respectively. For our final result we take the average among the $gri$ bands: $\Delta \overline{\text{HR}} = 0.031 \pm 0.018$, rms ratio $= 1.055 \pm 0.087$, and KS p-value is 0.07.

Here we perform several cross-checks. First, we repeat our analysis with apertures of different radii ranging from \SI{0.75}{\arcsecond} to \SI{2}{\arcsecond}, which yields results that agree with these values. Averaging over 6 different radii, $\Delta \overline{\text{HR}}$ increases by 0.004 corresponding to $\sim$25\% of the uncertainty. The largest $\Delta \overline{\text{HR}}$ shift is 0.008.
\vspace*{-3mm}
\begin{table}[h]
\begin{center}
\caption{Low-$z$ sample results}
\begin{tabular}{ |c|c|c|c| }
\hline
&&&\\[-2ex]
filter & $\Delta \overline{\text{HR}}$\tablenotemark{a} & rms ratio\tablenotemark{b} & KS p-value \\
\hline
&&&\\[-2ex]
$g$ & $0.029 \pm 0.018$ & $1.070 \pm 0.089$ & 0.08\\
$r$ & $0.037 \pm 0.018$ & $1.047 \pm 0.086$ & 0.04\\ 
$i$ & $0.028 \pm 0.018$ & $1.048 \pm 0.087$ & 0.10\\
\hline
&&&\\[-2ex]
avg & $0.031 \pm 0.018$ & $1.055 \pm 0.087$ & 0.07\\
\hline
\end{tabular}
\end{center}
$^{{\rm a}}$ $\Delta \overline{\text{HR}}$ = $\overline{\text{HR}}_{\text{faint}} - \overline{\text{HR}}_{\text{bright}}$\\
$^{{\rm b}}$ rms ratio = $\text{rms}_{\text{bright}}/ \text{rms}_{\text{faint}}$
\label{table:results}
\end{table}

As an additional test, we divide the sample of {\ntotsn} SNe~Ia into two sub-samples: 1) Legacy sample of {\ntotl} SNe~Ia from the Harvard-Smithsonian Center for Astrophysics surveys and the Carnegie Supernova Project, and 2) Foundation sample of {\ntotf} SNe~Ia from Pan-STARRS. The Legacy sample is older (2001-2010), heterogeneous, and includes galaxy-targeted surveys. The Foundation sample is more recent (2015-2017), homogeneous, better-characterized, and primarily follows SNe from surveys that do not target pre-selected galaxies. In principle, sample selection biases should not impact photometric measurements and here we test this assumption.

Table~3 shows the results of our correlation study for the Legacy and Foundation sub-samples. Both sub-samples show consistency between the bright and faint halves.
\vspace*{-3mm}
\begin{table}[h]
\begin{center}
\caption{Results for the Legacy and Foundation sub-samples}
\begin{tabular}{ |c|c|c|c|c| }
\hline
&&&&\\[-2ex]
sub-sample & filter & $\Delta \overline{\text{HR}}$ & rms ratio & KS p-value \\
\hline
&&&&\\[-2ex]
\multirow{3}{*}{Legacy} & $g$ & $0.023 \pm 0.028$ & $1.109 \pm 0.144$ & 0.27\\
& $r$ & $0.031 \pm 0.028$ & $1.075 \pm 0.139$ & 0.27\\ 
& $i$ & $0.037 \pm 0.028$ & $1.071 \pm 0.138$ & 0.27\\
\hline
&&&&\\[-2ex]
\multirow{3}{*}{Foundation} & $g$ & $0.029 \pm 0.023$ & $1.055 \pm 0.114$ & 0.13\\
& $r$ & $0.039 \pm 0.023$ & $1.057 \pm 0.114$ & 0.04\\ 
& $i$ & $0.038 \pm 0.022$ & $1.052 \pm 0.113$ & 0.05\\
\hline
\end{tabular}
\end{center}
\label{table:resultstwo}
\end{table}

We conclude that there is no statistically significant difference between the bright and faint half distributions, and therefore we find no evidence for SB-related bias in the low-redshift SN~Ia sample.

\section{Discussion and Conclusion}
\label{sect:conclusion}

We undertook an analysis of HR vs.\ SB with 292 low-$z$ SN~Ia from Foundation, the Harvard-Smithsonian Center for Astrophysics Surveys (CfA3, CfA4), and the Carnegie Supernova Project. 
We found no significant evidence for SB-related bias in this sample; the HR difference between the bright and faint SB subsets is $\Delta \overline{\text{HR}} = 0.031 \pm 0.018$. 
If such an HR difference turns out to be real, and it is not corrected in the analysis, this effect would cause a significant $w$-bias that is comparable in size to the largest systematic uncertainties in current analyses.

\begin{figure}[h]
    \begin{centering}
    \includegraphics[scale=0.55]{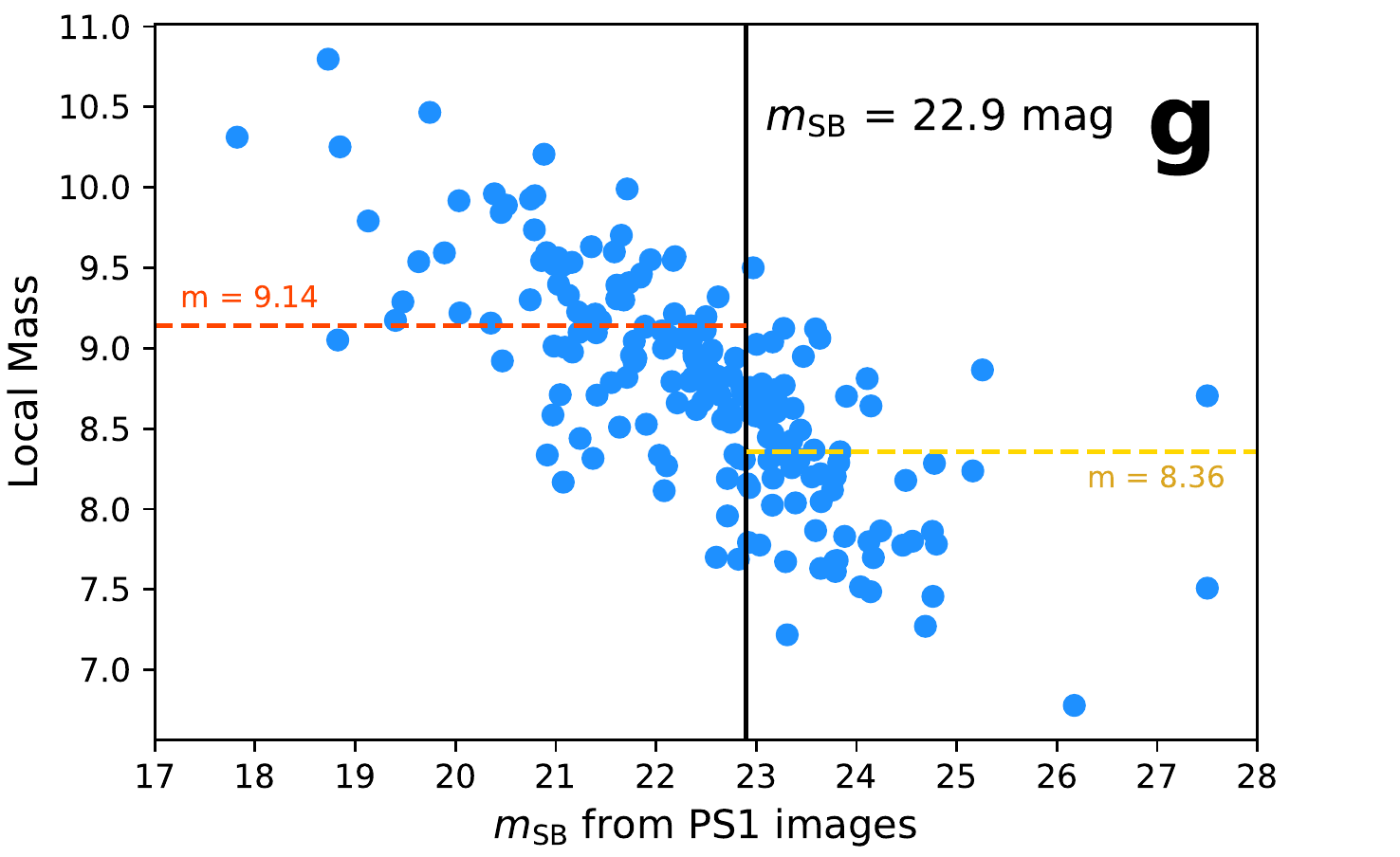}
    \caption{Correlation between our g-band SB and the local mass measurements from \citet{Jones18}. Vertical line indicates median $m_{\text{SB}}$ for the low-$z$ sample. Orange and yellow lines indicate median local mass for the bright and faint halves, respectively.}
    \label{fig:lcmasssb}
    \end{centering}
\end{figure}

It is also worth noting that SB-related biases could add unphysical correlations between SN luminosity and host galaxy properties. To illustrate this potential effect, Fig.~\ref{fig:lcmasssb} shows the correlation between our SB and the local mass measurements\footnote{\citet{Salim16} and \citet{San_Roman17} show that summing the results of pixel-by-pixel ``local" SED fitting give the same parameters as SED fitting to the global photometry.} from \citet{Jones18}, and shows that a non-zero $\Delta \overline{\text{HR}}$ could change the observed relationship between HR and local galaxy properties. The faint and bright subsets have a difference in median local mass of 0.8~dex. Note that there is a similar correlation between SB and local specific star formation rate.


With sufficiently large samples, we can gain further insight by studying $\Delta \overline{\text{HR}}$ as a function of galaxy morphology and color. Finally, the Vera C. Rubin Observatory\footnote{https://www.lsst.org/} is expected to produce at least an order of magnitude larger sample at low-redshift, which will reduce the $\Delta \overline{\text{HR}}$ uncertainty to well below the 1\% level in future cosmological analyses.

\acknowledgements
H.S.\ is supported by a Provost's scholarship at the University of Chicago.
This work was supported in part by the
Kavli Institute for Cosmological Physics at the
University of Chicago through an endowment from the
Kavli Foundation and its founder Fred Kavli.
R.K.\ is supported by DOE grant DE-SC0009924.
D.O.J.\ is supported by a Gordon and Betty Moore Foundation postdoctoral fellowship at the University of California, Santa Cruz and by NASA through the NASA Hubble Fellowship grant HF2-51462.001 awarded by the Space Telescope Science Institute, which is operated by the Association of Universities for Research in Astronomy, Inc., for NASA, under contract NAS5-26555.


\bibliographystyle{apj}
\bibliography{mybibfile}

\end{document}